\documentclass[%
 aps,prl,
 amsmath,amssymb,
preprint,%
]{revtex4-1}

\usepackage{graphicx}
\usepackage{dcolumn}
\usepackage{bm}
\usepackage{natbib}
\begin{document}

\title{Two-dimensional nanoscale imaging of gadolinium spins via scanning probe relaxometry with a single spin in diamond}

\author{M. Pelliccione} 
\author{B. A. Myers}
\author{L. Pascal}
\author{A. Das}
\author{A. C. Bleszynski-Jayich}
\affiliation{Department of Physics and the California NanoSystems Institute, University of California, Santa Barbara, Santa Barbara, CA 93106}

\date{\today}

\begin{abstract}
Spin-labeling of molecules with paramagnetic ions is an important approach for determining molecular structure, however current ensemble techniques lack the sensitivity to detect few isolated spins. In this Letter, we demonstrate two-dimensional nanoscale imaging of paramagnetic gadolinium compounds using scanning relaxometry of a single nitrogen vacancy (NV) center in diamond. Gadopentetate dimeglumine attached to an atomic force microscope tip is controllably interacted with and detected by the NV center, by virtue of the fact that the NV exhibits fast relaxation in the fluctuating magnetic field generated by electron spin flips in the gadolinium. Using this technique, we demonstrate a reduction in the $T_1$ relaxation time of the NV center by over two orders of magnitude, probed with a spatial resolution of 20 nm. Our result exhibits the viability of the technique for imaging individual spins attached to complex nanostructures or biomolecules, along with studying the magnetic dynamics of isolated spins. \end{abstract}

\maketitle

\section{\label{sec:introduction}INTRODUCTION}

Mapping the structure of biomolecules including proteins and nucleic acids is of significant importance, as the functionality of a biomolecule is directly related to its structure \citep{Hegyi1999}. For decades paramagnetic compounds, including gadolinium-based complexes such as Gd-DTPA, have been studied for their effect of reducing proton $\left(T_1\right)$ spin-lattice relaxation times \citep{Lauterbur1978}, making them widely used as nuclear magnetic resonance imaging (MRI) contrast agents \citep{Carr1985,Weinmann1984,Huang2013}. However, with conventional MRI techniques, the spatial resolution of contrast imaging is typically limited to the micron scale \citep{DiCorato2013,Kircher2012,Serres2012}. High-field electron paramagnetic resonance (EPR) has made possible nanometer-scale distance measurements between magnetically-interacting Gd$^{\rm 3+}$ spins tagged to proteins \citep{Potapov2010,Edwards2013}, but these approaches rely on a large ensemble of labeled molecules to obtain a measureable signal. A nanoscale scanning probe technique would enable non-averaged distance measurements on individual spin-labeled molecules, as well as an investigation of the magnetic dynamics of an isolated spin system. In this Letter, we demonstrate nanoscale imaging of Gd compounds using scanning relaxometry of a single nitrogen-vacancy (NV) center in diamond. The NV is an atomic-scale defect whose electronic spin, at ambient temperatures, exhibits several millisecond long longitudinal relaxation times, and can be optically polarized and interrogated on the single defect level \citep{Dobrovitski2013}. Together with its excellent photostability, biocompatibility, and noninvasiveness \citep{Kaufmann2013,Kucsko2013}, these properties make the NV a viable sensor for detecting and imaging individual spins in biological environments \citep{Rondin2014} and studying their fluctuation dynamics \citep{Cole2009,Schafer-Nolte2014}. 

Gadolinium (Gd$^{\rm 3+}$) ions are particularly interesting spin systems, as they have a large unpaired electron spin of $S = 7/2$ and fast spin dynamics in the GHz frequency range. These properties make Gd compounds particularly effective as MRI contrast agents \citep{Carr1984}, as the relaxation time of protons in water is reduced in the presence of Gd.  Analogously, the significant level of Gd spin noise at the frequency of the NV zero-field splitting (2.87 GHz) reduces the NV spin relaxation time $T_1$, depending on the proximity and concentration of Gd \citep{Steinert2013,Tetienne2013}. This susceptibility has inspired the technique of NV relaxometry to detect Gd spins. Few to single Gd spin sensitivities have been reported using single NV centers in nanodiamonds surrounded by a lipid bilayer \citep{Kaufmann2013} as well as with Gd compounds bonded to bulk diamond \citep{Sushkov2013}. Relaxometry with NV centers has also enabled the detection of ferromagnetic proteins in ambient conditions \citep{Ziem2013,Ermakova2013}. In addition, $T_1$-based imaging of Gd-tagged cellular structures has been demonstrated using ensembles of NV centers with a spatial resolution of 500 nm \citep{Steinert2013}, where the resolution is limited by the use of wide-field optical detection.

For molecular scale imaging, improvements in the spatial resolution and sensitivity of relaxometry measurements are necessary. This goal can be achieved by using NV-based scanning probe techniques. Scanning probes have already enabled nanoscale magnetic imaging using a variety of detection schemes; including static stray field imaging \citep{Maletinsky2012,Rondin2012,Rondin2013a,Haberle2013}, double electron-electron resonance \citep{Grinolds2013}, and proton magnetic resonance imaging \citep{Haberle2014,Rugar2014}. Relaxometry has the advantage of directly sensing electron spins, which have a magnetic moment nearly three orders of magnitude larger than protons spins sensed in NMR. In addition, relaxometry allows for the measurement of spins with $T_1$ times too short for DEER detection. Thus far, scanning $T_1$ relaxometry imaging has remained a challenge due to the requirement of stable, shallow NV centers with long $T_1$ times, coupled with the lengthy data acquisition times and associated scanning probe drift during the measurement. In this work, we overcome these challenges and perform two-dimensional NV relaxation imaging of a nanoscale volume of Gd electronic spins attached to an atomic force microscope (AFM) tip with 20 nm resolution. Furthermore, we show that with reasonable improvements this technique is capable of the sensitivity required to image a single isolated Gd spin. 

\section{METHODS}
\begin{figure}
\includegraphics[width = \textwidth]{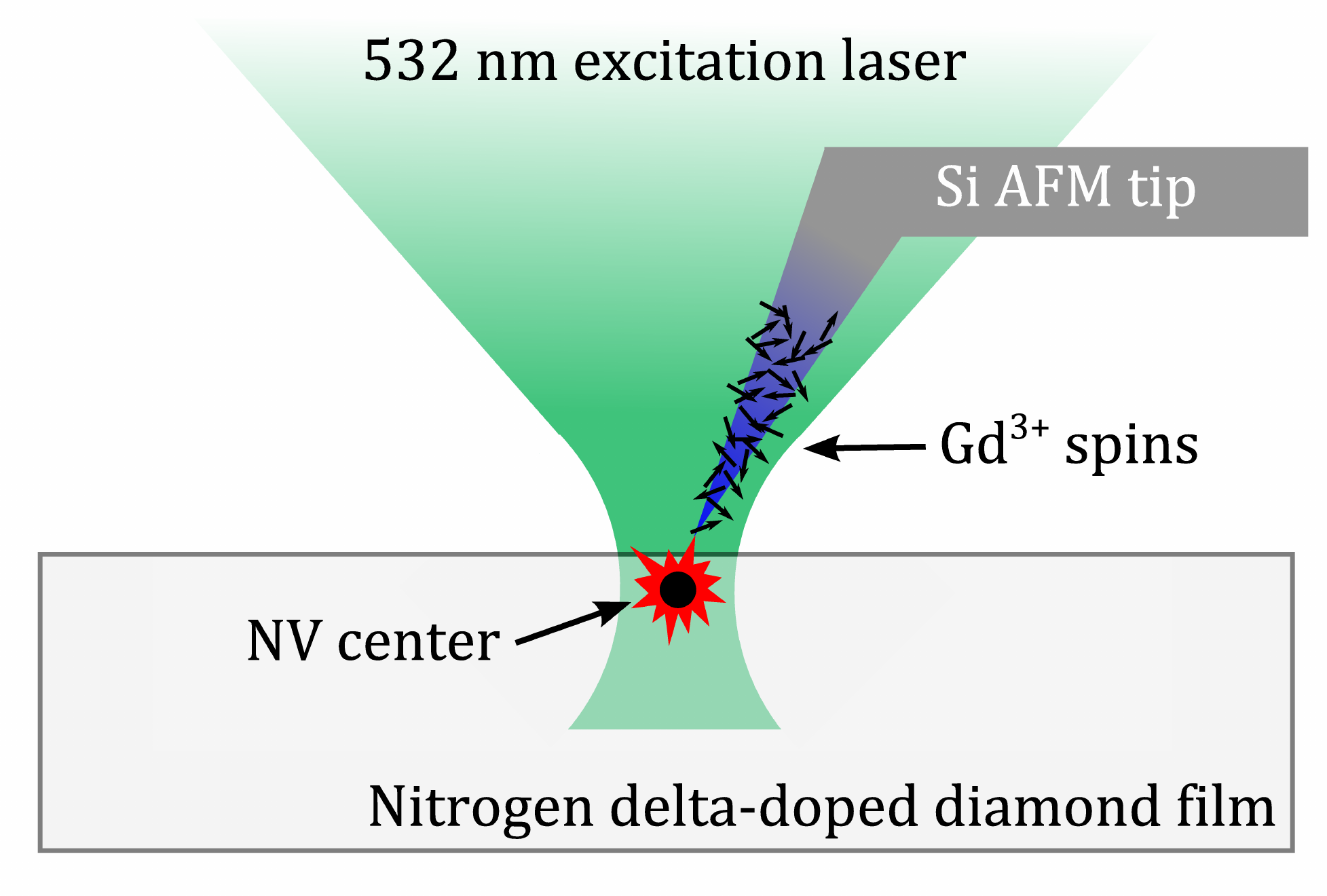}
\centering
\caption{Schematic of the scanning relaxometry measurement. A silicon AFM tip is coated with Gd compounds (Magnevist) and scanned near a shallow nitrogen vacancy center in single-crystal diamond. A confocal microscope excites and polarizes the NV spin with a laser power of 437 $\mu$W, and detects red photoluminescence to read out the NV spin polarization. This configuration allows for sensing a change in NV spin relaxation rate due to nearby Gd.}
\label{fig:scanningschem}
\end{figure}

As depicted schematically in Fig. \ref{fig:scanningschem}, the scanning probe setup combines a tuning fork-based AFM, a top-down confocal microscope, and bulk diamond containing NV centers near the surface. Gadolinium is attached to a silicon AFM tip by submerging the cantilever in a chelated Gd solution (Gadopentetate dimeglumine in water (Magnevist), concentration 30 mM) for several minutes. Experiments are performed at ambient conditions and in the absence of an applied static magnetic field. During the measurement, the AFM is operated in tapping mode with a tapping amplitude of 1 nm RMS. The detection scheme is all-optical; the NV center is polarized into the $\left| m_s = 0 \right>$ state of the ground state triplet with a non-resonant green laser pulse, and read out via spin-dependent photoluminescence during a subsequent laser pulse a time $\tau$ later, \citep{Oort1988} as depicted in Fig. \ref{fig:bulkt1}(a). During the dark time $\tau$, the NV polarization relaxes to an equilibrium mixed state of $\left| 0 \right>$, $\left| +1 \right>$ and $\left| -1 \right>$ with a characteristic time of $T_1$. In the presence of Gd  a distance $r$ from the NV, the NV $T_1$ is reduced according to the expression
\begin{equation}
T_1^{-1} = T_{1, \rm int}^{-1} + \Gamma_{\rm Gd}(r),
\end{equation}
where $T_{1, \rm int}^{-1}$ is the intrinsic NV relaxation rate in the absence of Gd and $\Gamma_{\rm Gd}(r)$ is the additional relaxation rate due to Gd. The NV is relaxed by magnetic fields perpendicular to its symmetry axis that appear static in the rotating frame, or equivalently that oscillate at the Larmor frequency $\omega_{\rm NV}/(2\pi)  \approx 2.87$ GHz in the lab frame. Gadolinium has a magnetic noise spectral density that is broadened into the GHz range \citep{Steinert2013}, and hence for sufficiently small $r$ and sufficiently long $T_{1, \rm int}$, $\Gamma_{\rm Gd}$ can be of the same order or larger than $T_{1, \rm int}^{-1}$, leading to a detectable change in the NV $T_1$. 

The diamond film used in this work was grown with a nitrogen delta-doping method, in which nitrogen is incorporated into the diamond sample during epitaxial growth \citep{Ohno2012,Ohashi2013}. Growth was followed by electron irradiation and annealing for vacancy creation and diffusion \citep{Ohno2012}. The NV centers used in the present work are $8-10$ nm below the diamond surface, as determined by magnetic resonance depth imaging \citep{Myers2014}, and typical measured $T_{1, \rm int}$ times are about $1-4$ ms at room temperature. 

\section{Gadolinium-NV RELAXOMETRY}
\begin{figure}
\includegraphics[width = \textwidth]{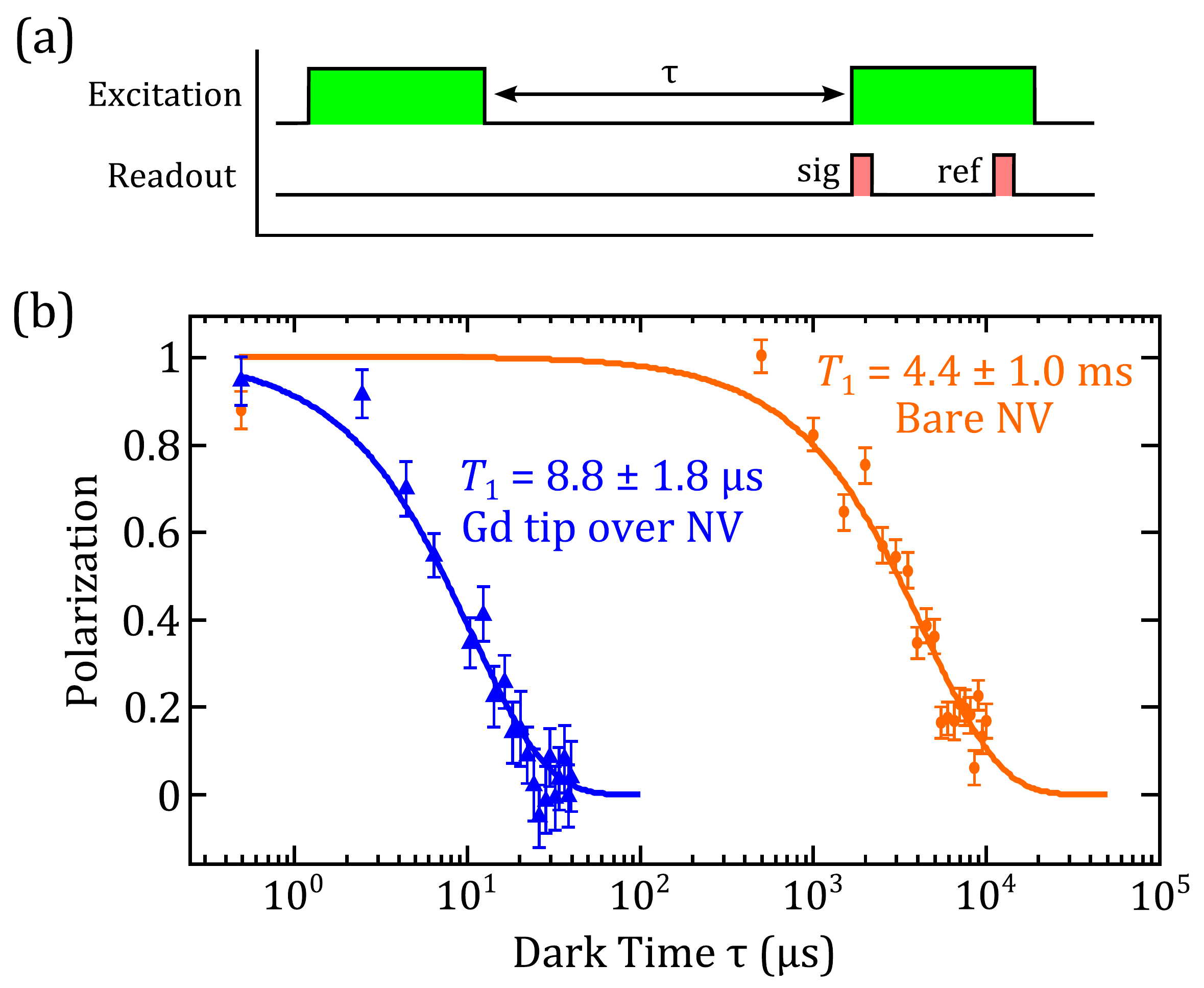}
\centering
\caption{(a) Pulse sequence used to measure the $T_1$ relaxation time, where the dark time $\tau$ is varied. Each signal pulse has a duration of 350 ns, and is followed by a reference pulse after a delay of 2.5 $\mu$s. The dark time $\tau$ does not include a 500 ns dark time after the initialization pulse, which allows for full depopulation of the metastable state. (b) Measurement of spin relaxation of a single NV center with the Gd-coated AFM tip positioned over the NV center (blue triangles) and moved 5 $\mu$m away (orange circles). The vertical axis is plotted in terms of NV polarization, with a polarization of 1 referring to the NV in the $\left| 0 \right>$ state, and a polarization of 0 referring to the NV in an equilibrium mixed state of $\left| 0 \right>$, $\left| +1 \right>$ and $\left| -1 \right>$. The $T_1$ times are extracted by fits (solid lines) of the data to an exponential decay with decay constant $T_1$.}
\label{fig:bulkt1}
\end{figure}

We first show that positioning the Gd-coated tip in close proximity to a shallow NV center can reproducibly change its relaxation time. Figure \ref{fig:bulkt1}(b) shows the $T_1$ relaxation curves for a single NV at two tip positions; centered above the NV center (blue) and 5 $\mu$m laterally displaced from the NV (orange). At a tip-NV separation of 5 $\mu$m, the tip is sufficiently far away such that $T_1 = T_{1,\rm int}$. By positioning the tip within tens of nanometers of the NV, we observe an almost three orders of magnitude reduction in $T_1$, from 4.4 ms to 8.8 $\mu$s. This measurement can be cycled with consistent results, which provides verification that the surface is not becoming contaminated with Gd, and that the tip retains its integrity; both are critical requirements for faithful imaging.

The measurement of the full relaxation curve shown in Fig. \ref{fig:bulkt1}(b) can take many hours, limited mainly by photon shot noise, which is impractical for imaging experiments. Data acquisition time is of heightened importance for two- or three-dimensional imaging, as the number of data points scales rapidly as the spatial resolution is increased. Furthermore, it is difficult to keep the tip-NV separation stable in ambient conditions with traditional AFM techniques over these time scales, due mainly to thermal drift. To reduce data acquisition time and mitigate measurement errors induced by thermal drift, we sample only a small subset of $\tau$ points on the curve in Fig. \ref{fig:bulkt1}(b) when imaging. The set of $\tau$ points we use is judiciously chosen to maximize the signal-to-noise ratio (SNR). It is straightforward to show that the SNR is maximized for a fixed $\tau$ approximately equal to $T_1$. However, when performing scanning measurements, $T_1$ can vary across the sample by many orders of magnitude, and hence different $\tau$ values optimize the SNR at different positions in the scan. 

\begin{figure}
\includegraphics[width = \textwidth]{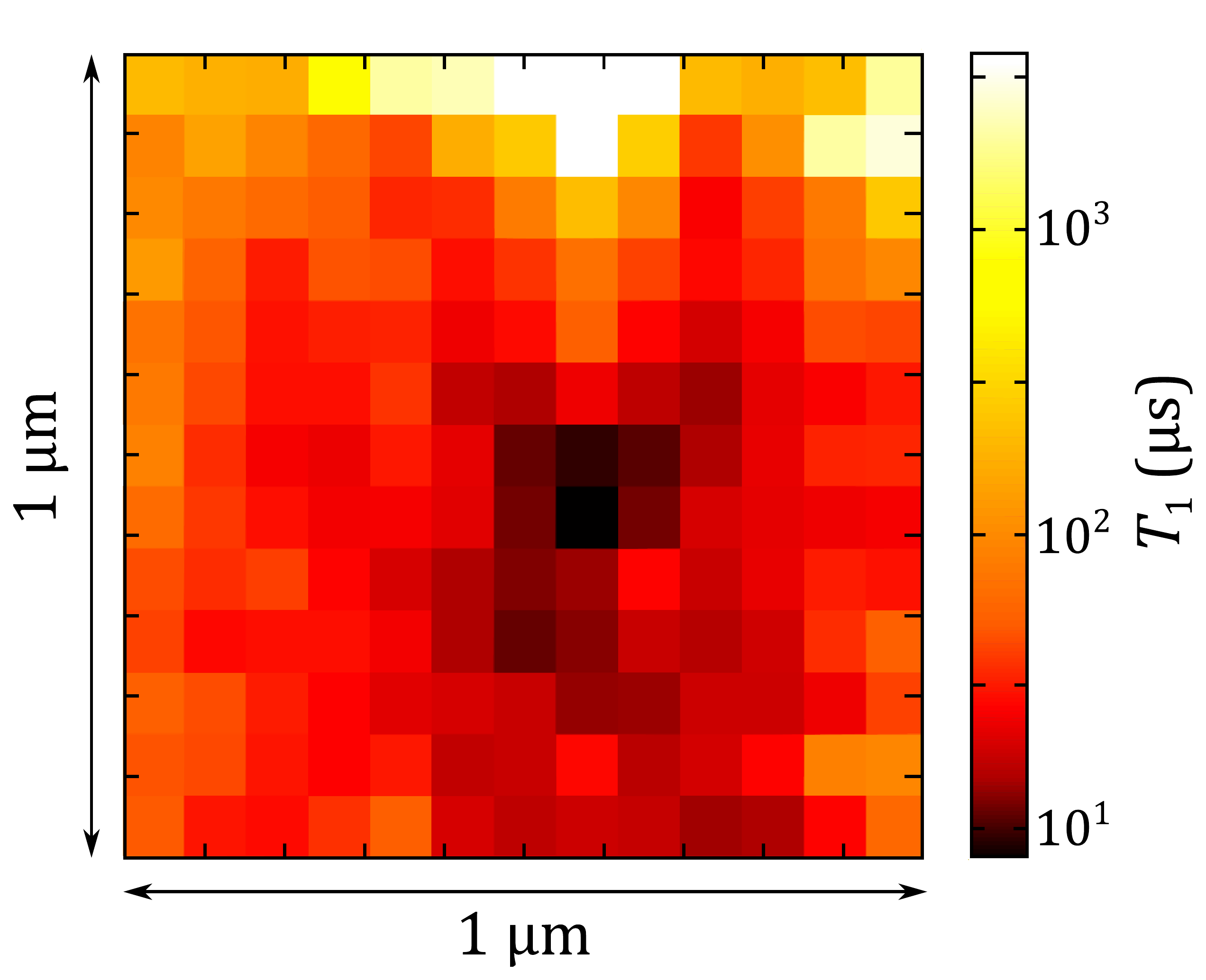}
\centering
\caption{Two-dimensional map of the $T_1$ relaxation time of the NV center versus Gd-coated tip position. The $T_1$ times were inferred from fixed $\tau$ measurements with $\tau = \left(4, 8, 40, 80, 400, 800\right)$ $\mu$s at each pixel. The distinct reduction in $T_1$ in the center of the image indicates the closest approach of the tip to the NV.}
\label{fig:t1map}
\end{figure}

A two-dimensional map of the NV $T_1$ versus tip position is presented in Fig. \ref{fig:t1map}. There is a clear and highly localized reduction in $T_1$ near the center of the scan that indicates the location of closest approach between the Gd-coated tip and the NV center. As expected, the $T_1$ increases as the tip-NV separation is increased until the original $T_{1, \rm int}$ of the NV center is observed along the periphery of the scan area. The observation of $T_{1, \rm int}$ is important because it indicates that there is no significant Gd contamination on the diamond surface during the scan. To mitigate thermal drift, observed to be approximately 1 nm/min, after each pixel an image registration algorithm \citep{Guizar-Sicairos2008} was used to realign the tip with the NV center. The alignment image was provided by the near field optical profile of the NV photoluminescence (PL) in the presence of the tip, which allowed for a reproducible alignment with a maximum error of 10 nm. 

The image in Fig. \ref{fig:t1map} is compiled from a set of measurements with $\tau = \left(4, 8, 40, 80, 400, 800\right)$ $\mu$s, which span the range of $T_1$ times accessible in the scan area. To generate a $T_1$ image from the fixed $\tau$ measurements, a fit to an exponential decay is performed for the data taken at each fixed $\tau$. The $T_1$ times extracted from the six fixed $\tau$ measurements are then averaged, weighted by the error in their respective fit, to arrive at a final $T_1$ time that is plotted in the image. The $T_1$ exponential fit is complicated by the dependence of the measured PL on tip position, due to a combination of shadowing and near field effects from the tip, and a reduced PL when $T_1$ becomes comparable to the sub-$\mu$s metastable state relaxation time \citep{Tetienne2013}. Therefore, at each tip position the steady state PL under laser excitation is measured and included in the $T_1$ fit. 

\begin{figure}
\includegraphics[width = 0.9\textwidth]{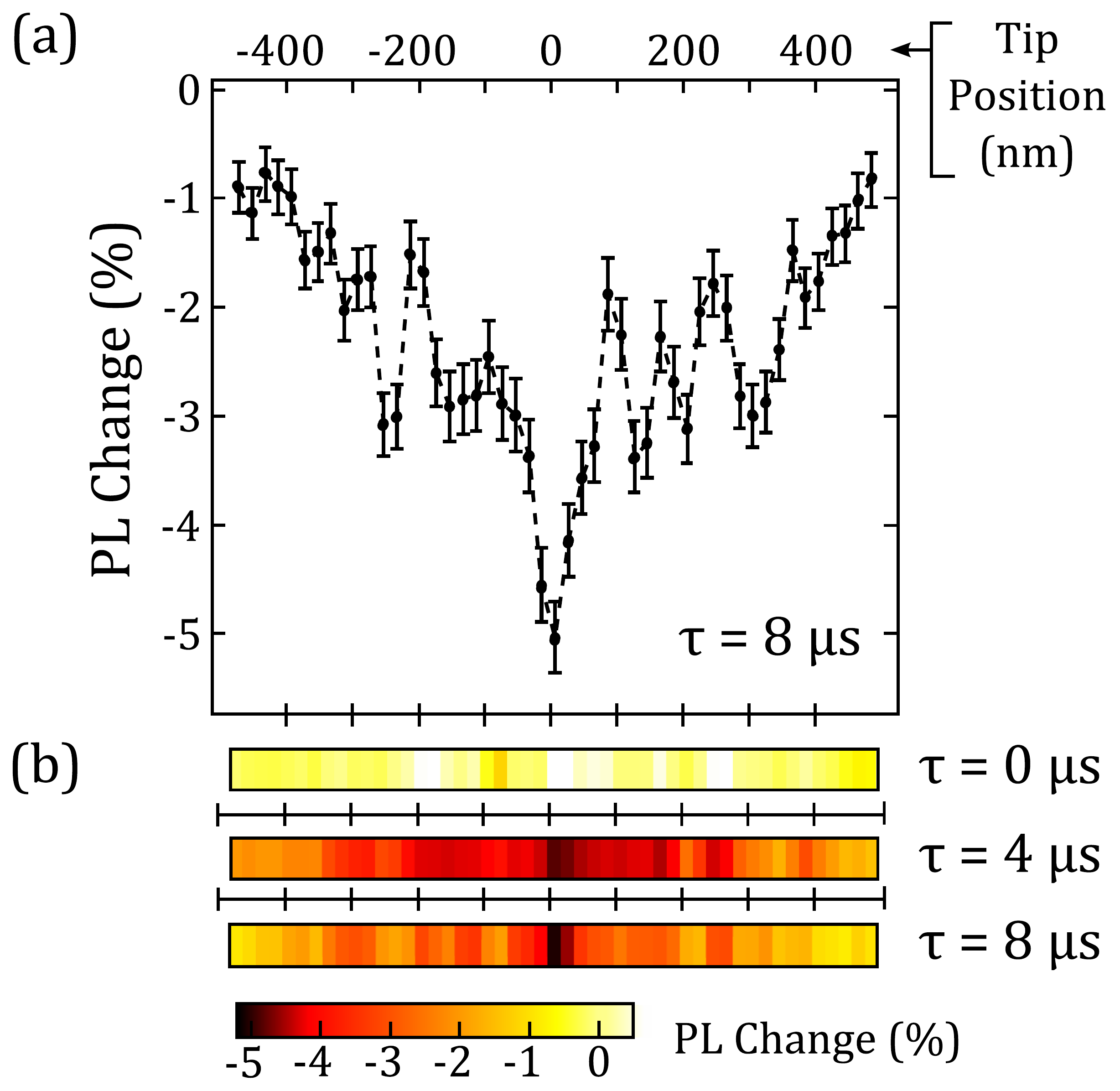}
\centering
\caption{(a) One-dimensional line cut of the data in Fig. \ref{fig:t1map}, showing the measured PL change for a fixed dark time $\tau = 8$ $\mu$s. The error bars are computed from two consecutive line scans with nearest-neighbor averaging, and are attributed to photon shot noise and tip drift. (b) Line scans for $\tau = \left(0, 4, 8\right)$ $\mu$s. There is no discernable contrast at $\tau = 0$ $\mu$s, but at longer $\tau$ there is a clear reduction in $T_1$ at the center of the line scan that depends sharply on tip position. The total data acquisition time at each tip position was 6 minutes.}
\label{fig:fixedtmeas}
\end{figure}

Figure \ref{fig:fixedtmeas}(a) shows a one-dimensional line cut of a single $\tau = 8$ $\mu$s measurement in Fig. \ref{fig:t1map} taken through the location of the NV center.  In order to maximize the SNR when the tip is near the NV, we chose this fixed $\tau$ value to be around 8.8 $\mu$s, the expected $T_1$ when the tip is near the NV from Fig. \ref{fig:bulkt1}(b). Plotted on the vertical axis is the percentage change in PL at each tip position. This change, defined as $\left[\mbox{PL(sig)}/\mbox{PL(ref)} -1\right]$ and heretofore referred to as contrast, is equal to zero if the NV is polarized in the $\left| 0 \right>$ state, and becomes negative as the NV evolves into an unpolarized state. The one-dimensional color plots in Fig. \ref{fig:fixedtmeas}(b) show how the fixed $\tau$ contrast changes with choice of $\tau$. At $\tau = 0$ $\mu$s, the state of the NV is polarized at all tip positions, resulting in little contrast across the entire scan. Increasing $\tau$ to 4 $\mu$s begins to reveal contrast in the center of the line scan where the NV $T_1$ is the shortest, while the contrast remains near zero at the extremes of the line scan where $T_1$ is the longest. The contrast at the center of the line scan is further enhanced at $\tau = 8$ $\mu$s, where $T_1\approx\tau$.  

Figure \ref{fig:t1sim} plots $T_1$ as a function of tip position zoomed in to a 300 nm wide region in the center of Fig. \ref{fig:fixedtmeas}. In this case, $T_1$ is extracted from a fit to the $\tau=\left(4, 8\right)$ $\mu$s data shown in Fig. \ref{fig:fixedtmeas}, as these fixed $\tau$ points provide the best estimate for $T_1$ in this range. The data show an approximately 50 nm wide feature and importantly, from the slope of the feature edges, a spatial resolution estimated to be 20 nm. This spatial resolution is set by the 20 nm scan step size, which was chosen to be slightly larger than the combined effect of $\sim$10 nm AFM drift and $\sim$10 nm repeatability of the image registration algorithm used per point. Pushing to higher spatial resolution will require first an improvement in the AFM drift during the measurement, and eventually shallower NV centers. 

\section{SIMULATIONS}
\begin{figure}
\includegraphics[width = 0.7\textwidth]{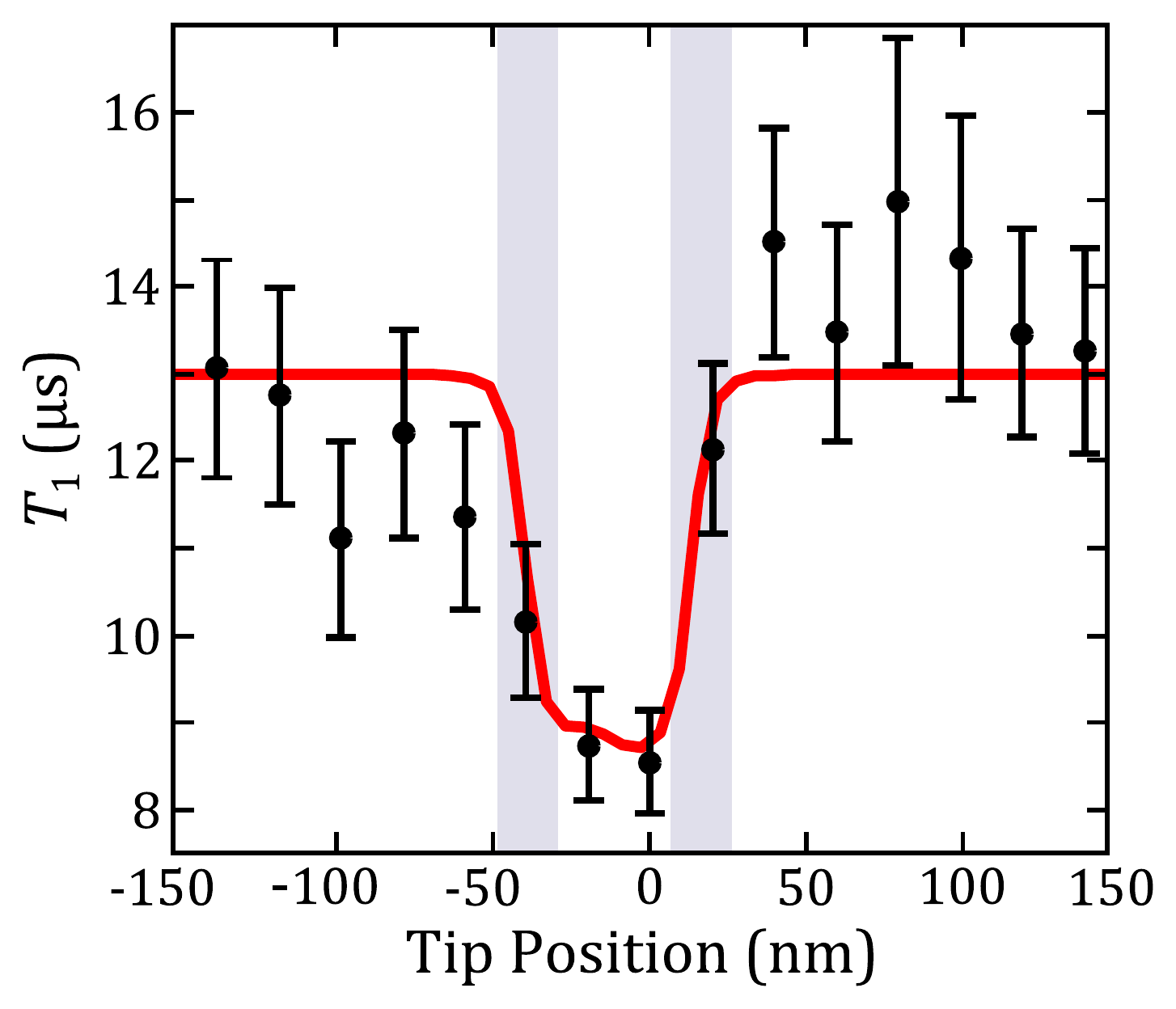}
\centering
\caption{Extracted $T_1$ times from the data presented in Fig. \ref{fig:fixedtmeas}. A spatial resolution of 20 nm is deduced from the change in $T_1$ versus tip position in comparison to the magnitude of the vertical error bars. The shaded blue regions are 20 nm wide as a guide to the eye. The data is modeled by a simulation that computes, as a function of scan position, the magnetic field from a finite 50 nm $\times$ 50 nm surface of Gd. The result (red solid line) indicates a Gd density of 10.7 spins/nm$^2$ for an NV depth of 10 nm.}
\label{fig:t1sim}
\end{figure}

The 50 nm wide plateau of reduced $T_1$ in the center of Fig. \ref{fig:t1sim} represents a region of the tip with locally enhanced Gd concentration over a background, the densities of which can be estimated from a simulation shown in the red trace. The model for our simulation places a two-dimensional layer of Gd ions on a surface and computes the magnetic field the NV would experience from the ensemble of Gd spins at each scan position \citep{Kaufmann2013}. Assuming the Gd samples all $\left\{\left|m_s\right>\right\}$ in the $S=7/2$ Hilbert space in a thermal mixture, we compute the mean square perpendicular magnetic field the NV experiences from each Gd spin, $\left<\left[\mathbf{B}_\perp\left(\mathbf{r}\right)\right]^2\right>$. In this expression, $\left<\ldots \right>$ denotes a mean square average over the $\left\{\left|m_s\right>\right\}$ subspace taken by a trace over the density matrix of the mixed state. The magnetic field from a single Gd spin is given by
\begin{equation}
\mathbf{B}\left(\mathbf{r}\right)= \frac{\mu_0}{4\pi}\frac{g_{\rm Gd}\mu_B}{\left|\mathbf{r}\right|^3}\left[\mathbf{S} - \frac{3\mathbf{r}\left(\mathbf{r}\cdot\mathbf{S}\right)}{\left|\mathbf{r}\right|^2}\right],
\end{equation}
where $\mathbf{S}$ is the Gd electron spin vector, $g_{\rm Gd} = 2$ is the Gd electron g-factor, and $\mu_B$ is the Bohr magneton. The NV relaxation rate due to the fluctuating field of a single Gd spin at position $\mathbf{r}$ is given by
\begin{equation}
\Gamma_{\rm Gd}\left(\mathbf{r}\right) = \frac{3\tau_c\gamma_{\rm NV}^2}{1+\omega_{\rm NV}^2\tau_c^2} \left<\left[\mathbf{B}_{\perp}\left(\mathbf{r}\right)\right]^2\right>,
\end{equation}
where $\gamma_{\rm NV} = 2\pi \times 28$ GHz/T is the NV electron spin gyromagnetic ratio, and $\tau_c$ is the effective correlation time of the Gd noise spectrum, taken to be 0.36 ns \citep{Kim2009}. With a 10 nm deep NV, the simulation result in Fig. \ref{fig:t1sim} predicts a background concentration of 7.2 spins/nm$^2$, with an additional 3.5 spins/nm$^2$ in a 50 $\times$ 50 nm region of the tip, which reduces $T_1$ from 13 $\mu$s to 8.8 $\mu$s at the center of the line scan. 

\begin{figure}
\includegraphics[width = 0.7\textwidth]{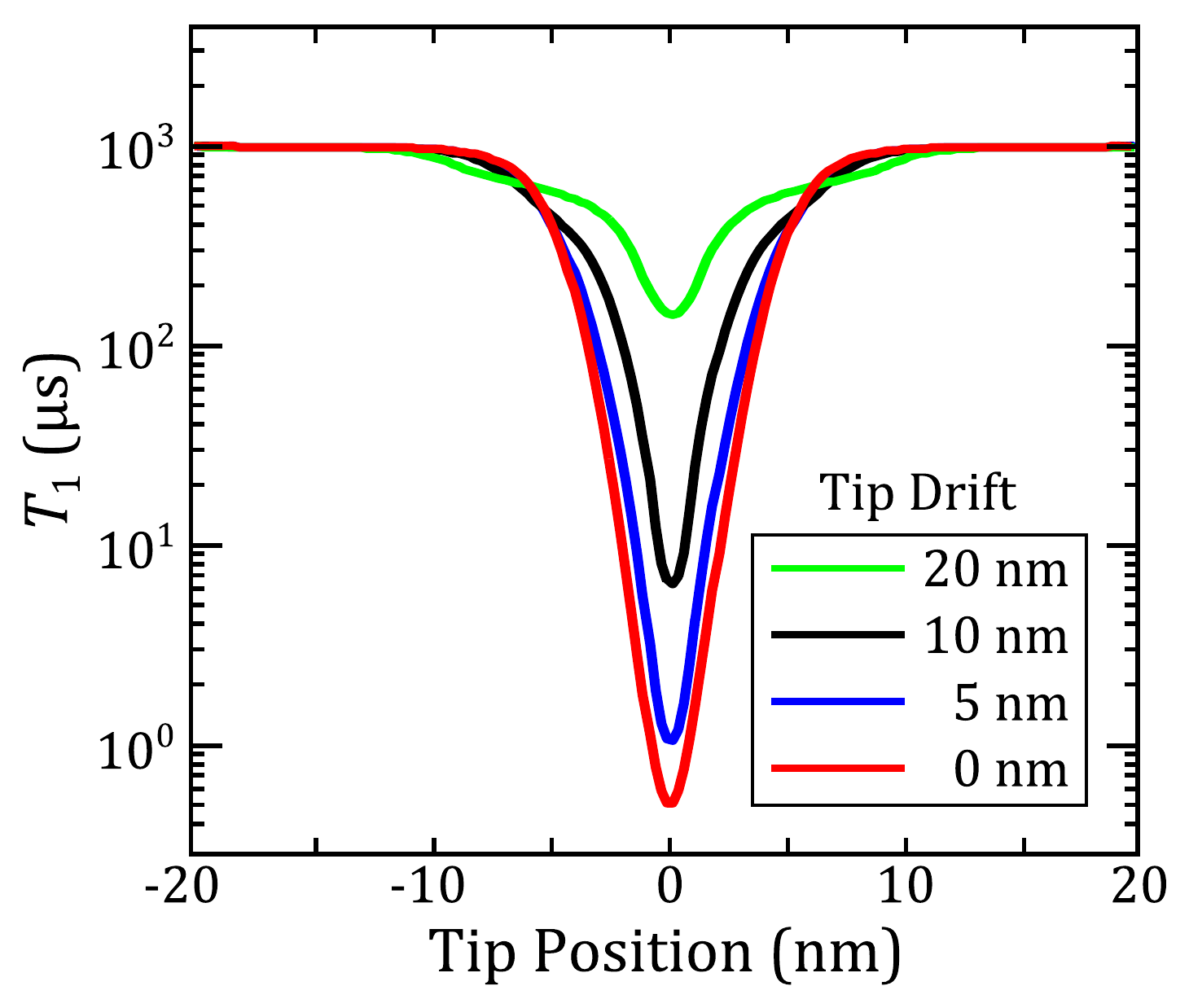}
\centering
\caption{Simulated $T_1$ response to a single Gd spin for a 3 nm deep NV center with $T_{1,\rm int}= 1$ ms, accounting for various magnitudes of tip drift during the measurement. The minimum $T_1$ time observed depends strongly on the size of the drift, increasing from 0.50 $\mu$s with no drift to $\left(1.04, 6.36, 145\right)$ $\mu$s with $\left(5, 10, 20\right)$ nm of drift, respectively.}
\label{fig:t1drift}
\end{figure}

We now turn to a more detailed discussion of spatial resolution and sensitivity, which are intimately related in this experiment. In particular, we focus on the goal of imaging a single Gd spin. Thermal AFM tip drift during the measurement can have dramatic effects, as is evidenced by the simulation results shown in Fig. \ref{fig:t1drift}. Plotted are one-dimensional $T_1$ images simulated with different magnitudes of tip drift for a single Gd target spin, a 3 nm deep NV center and $T_{1, \rm int} = 1$ ms. To calculate these results, first the dependence of $T_1$ on tip position is calculated in the absence of drift, shown as the red trace with the largest dip in Fig. \ref{fig:t1drift}. When drift is encountered experimentally, the result will be a sampling of many tip positions during the measurement, which serves to blur out the effect of the Gd spin. This can be modeled by taking a sampling of $T_1$ times around each tip position, with a spatial width equal to the magnitude of the drift during the acquisition time for each measurement point. We can then sum the exponential decay curves of each $T_1$ time in the sampling area to represent the curve that would be measured experimentally. Although this curve is a sum of exponential decays with different time constants, we can do a least squares fit to a single exponential decay with one time constant to obtain an averaged $T_1$ response. Carrying out this procedure for different drift magnitudes of $\left(5, 10, 20\right)$ nm shows a stark reduction in the predicted $T_1$ response as drift is increased, evident on a log scale. Notably, the shortest measured $T_1$ time decreases from 145 $\mu$s with 20 nm of drift to 0.50 $\mu$s with no drift. Using the current experimental parameters of a 10 nm deep NV, $T_{1, \rm int} = 4.4$ ms, 10 nm of drift per measurement point, and 70 kCounts/second of photon counts from the NV center, we can use the simulation results to predict a single spin sensitivity that accounts for thermal drift. Under these conditions with a single Gd spin, one predicts a minimum $T_1$ time of 715 $\mu$s, and a SNR of 1 to be reached in 30 seconds of averaging time. 

To reach the goal of imaging single Gd spins, we address several areas for improvement. Using shallower NV centers would provide a much larger signal, since $\Gamma_{\rm Gd}(r) \sim r^{-6}$ for a single Gd spin. Thermal drift, estimated to be about 1 nm/min in this work, can be reduced by using active drift compensation at the expense of measurement complexity, which has been shown to improve drift to 5 pm/min in ambient conditions \citep{King2009}. Operating at cryogenic temperatures can also reduce drift significantly, but also at the expense of measurement complexity and incompatibility with biological systems. Photon shot noise can be improved by increasing the number of counts from the NV center. In the current geometry, the AFM tip and objective are on the same side of the diamond sample, which leads to partial shadowing of the NV center by the AFM tip. Using a geometry where the objective and AFM tip are on opposite sides of the sample \citep{Myers2014} would improve the NV count rate, as well as allow for the use of an oil immersion objective with a large numerical aperture. Alternatively, collection efficiency can be dramatically improved by structuring the diamond with nanopillars, in particular by using an NV diamond nanopillar as the scanning probe \citep{Maletinsky2012,Grinolds2013}, though presently NV spin properties in such nanostructures are poor compared to those in bulk diamond. 

We now estimate the sensitivity to a single Gd spin for optimized conditions of reduced drift, higher photon counts, and shallower NV centers based on the discussions above. Using a 3 nm deep NV, $T_{1,\rm int} = 1$ ms to account for shorter $T_1$ times typically seen in shallower NV centers \citep{Myers2014}, 1 nm of drift per measurement point, and 120 kCounts/second from the NV center, one predicts a minimum $T_1$ time of 0.55 $\mu$s, and a SNR of 1 to be reached in 10 ms of averaging time. We note that these improvements are realistic: $2-3$ nm deep NV centers have already been demonstrated as external nuclear spin sensors \citep{Loretz2014,Muller2014}, and oil immersion objectives readily achieve such photon count rates. This result demonstrates the feasibility of performing scanning relaxometry with NV centers, and provides a roadmap for controllably detecting single Gd electron spins with scanning probe microscopy.

\begin{acknowledgments}
The authors thank K. Ohno and D. D. Awschalom for diamond growth instruction. This work was supported by the DARPA QuASAR program, the Air Force Office of Scientific Research under Award No. FA9550-13-1-0198, and the MRSEC Program of the National Science Foundation under Award No. DMR 1121053. B. A. M. is supported through the Department of Defense (NDSEG). 
\end{acknowledgments}

\newpage

\end{document}